\documentclass[
	secnumarabic,
	amssymb,
	amsmath,
	nofootinbib,
	nobibnotes, 
	aps,
	12pt,
	superscriptaddress 
	]
	{revtex4-2}
	
\usepackage{graphics}
\usepackage{graphicx}
\usepackage{color}
\usepackage{comment}
\usepackage{amsmath}
\usepackage{bm}

\begin{document}

%
%
\title{
New determination of the branching ratio of the structure dependent radiative  $K^{+} \to e^{+} \nu_{e} \gamma$ }

%
%
\author{A.~Kobayashi}
\affiliation{Department of Physics, Chiba University, Chiba, 263-8522, Japan}

\author{H.~Ito}
\thanks{
Present address:Department of Physics, Tokyo University of Science, Noda, Chiba, 278-8510, Japan.  
}
\affiliation{Department of Physics, Chiba University, Chiba, 263-8522, Japan}

\author{S.~Bianchin}
\affiliation{TRIUMF, Vancouver, BC, V6T 2A3, Canada}

\author{T.~Cao}
\affiliation{Physics Department, Hampton University, VA 23668, USA}

\author{C.~Djalali}
\affiliation
{Department of Physics and Astronomy, University of Iowa, Iowa City, IA 52242, USA} 

\author{D.H.~Dongwi}
\affiliation{Physics Department, Hampton University, VA 23668, USA}

\author{T.~Gautam}
\affiliation{Physics Department, Hampton University, VA 23668, USA}

\author{D.~Gill}
\affiliation{TRIUMF, Vancouver, BC, V6T 2A3, Canada}

\author{M.D.~Hasinoff}
\affiliation{Department of Physics and Astronomy, University of British Columbia, Vancouver, BC, V6T 1Z1, Canada}

\author{K.~Horie}
\affiliation{Department of Physics, Osaka University, Osaka, 560-0043, Japan}

\author{Y.~Igarashi}
\affiliation{High Energy Accelerator Research Organization (KEK), Tsukuba, 305-0801, Japan}

\author{J.~Imazato}
\affiliation{High Energy Accelerator Research Organization (KEK), Tsukuba, 305-0801, Japan}

\author{N.~Kalantarians}
\thanks{
Present address:Department of Natural Sciences, Virginia Union University, Richmond, VA 23220, USA.}
\affiliation{Physics Department, Hampton University, VA 23668, USA}

\author{H.~Kawai}
\affiliation{Department of Physics, Chiba University, Chiba, 263-8522, Japan}

\author{S.~Kimura}
\affiliation{Department of Physics, Chiba University, Chiba, 263-8522, Japan}

\author{S.~Kodama}
\affiliation{Department of Physics, Chiba University, Chiba, 263-8522, Japan}

\author{M.~Kohl}
\affiliation{Physics Department, Hampton University, VA 23668, USA}

\author{H.~Lu}
\affiliation{Department of Physics and Astronomy, University of Iowa, Iowa City, IA 52242, USA}

\author{O.~Mineev}
\affiliation{Institute for Nuclear Research, Moscow, 117312, Russia}

\author{P.~Monaghan}
\thanks{
Present address:Department of Physics, Christopher Newport University, Newport News, VA 23606, USA.  
}
\affiliation{Physics Department, Hampton University, VA 23668, USA}

\author{S.~Shimizu}
\thanks{
Corresponding author:
suguru@phys.sci.osaka-u.ac.jp}
\affiliation{Department of Physics, Osaka University, Osaka, 560-0043, Japan}


\author{M.~Tabata}
\affiliation{Department of Physics, Chiba University, Chiba, 263-8522, Japan}

\author{R.~Tanuma}
\thanks{Deceased}
\affiliation{Department of Physics, Rikkyo University, Toshima, 171-8501, Japan}

\author{A.~Toyoda}
\affiliation{High Energy Accelerator Research Organization (KEK), Tsukuba, 305-0801, Japan}

\author{H.~Yamazaki}
\affiliation{High Energy Accelerator Research Organization (KEK), Tsukuba, 305-0801, Japan}

\author{N.~Yershov}
\affiliation{Institute for Nuclear Research, Moscow, 117312, Russia}

\collaboration{J-PARC E36 Collaboration}

%
%
\begin{abstract}
The branching ratio of the structure dependent (SD) radiative $K^{+} \to e^{+} \nu_{e} \gamma$ decay relative to that of the $K^+\rightarrow e^+ \nu_{e} (\gamma)$ decay including the internal bremsstrahlung (IB) process ($K_{e2(\gamma)}$) has been measured in the J-PARC E36 experiment using plastic scintillator/lead sandwich \color{black} detectors, \color{black} in contrast to the previous E36 measurement, which used a CsI(Tl) calorimeter.  \color{black}
In the analysis, the effect of IB was also taken into account in the SD radiative decay as $K_{e2\gamma(\gamma)}^{\rm SD}$.
By combining the new data with \color{black} the previous E36 result  \color{black} after revision for the IB correction for $K_{e2\gamma(\gamma)}^{\rm SD}$, a new value $Br(K_{e2\gamma(\gamma)}^{\rm SD})/Br(K_{e2(\gamma)})=1.20\pm0.07$ has been determined.
This is consistent with a recent lattice QCD calculation, but larger than the expectation of Chiral Perturbation Theory (ChPT) at order $O(p^4)$ and the previous KLOE value. 
Using the method to relate form factor and branching ratio described in the KLOE paper, the present result is also consistent with the form factor prediction based on a gauged nonlocal chiral quark model, but larger than \color{black} that from \color{black} ChPT at order $O(p^6)$. 
\end{abstract} 
\maketitle

\section{Introduction}
Semi-leptonic radiative decays of $K$-mesons, $K^+\rightarrow l \nu \gamma$ ($K_{l2\gamma}$), provide an excellent testing ground for hadron structure models making use of low-energy effective Lagrangians inspired by Chiral Perturbation Theory (ChPT). 
It is expected that branching ratio measurements can provide simple but excellent constraints on models. 
The radiative decays of mesons normally consist of an internal bremsstrahlung (IB) process,  a hadronic structure-dependent (SD) process, \color{black} and interference terms between the IB and SD parts of the decay amplitude~\cite{biji1, biji2, rev_mod}.  \color{black}
The branching ratio of $K^+ \rightarrow e^+ \nu_{e}$ ($K_{e2}$) is known to be strongly suppressed down to $\sim O(10^{-5})$ due to the helicity suppression of the weak charged current. 
On the other hand, the SD radiative process ($K_{e2\gamma}^{\rm SD}$) is not subject to this helicity suppression, and its branching ratio, $Br(K_{e2 \gamma}^{\rm SD})$, is comparable to that of $K_{e2}$. 
This is an essential characteristic of the $K_{e2 \gamma}^{\rm SD}$ decay, which can be measured  with less background compared to the SD radiative $K^+\rightarrow \mu^+ \nu_{\mu} \gamma$ ($K_{\mu 2 \gamma}$) decay. 
\color{black} Also, it should be emphasized that the interference terms are negligible for the $K_{e2\gamma}^{\rm SD}$ decay, but they are important in the $K_{\mu 2 \gamma}$ decay. \color{black}
Therefore, the measurement of the $K_{e2 \gamma}^{\rm SD}$ decay can provide \color{black} information about various hadron structure models.  \color{black}

On the other hand, the $K_{e2\gamma}^{\rm SD}$ decay is a dominant background in the experiment to search for lepton universality violation by measuring the ratio of the $K_{e2}$ and $K^+ \rightarrow \mu^+ \nu_{\mu}$ ($K_{\mu 2}$) branching ratios ($R_K$)~\cite{RK_theo1,RK_theo2,RK_theo3,RK_theo4}.
Because the hadronic decay constants for the two decays are common, they are cancelled out in the $R_K$ calculation. 
The Standard Model (SM) prediction of $R_K=(2.477\pm 0.001)\times 10^{-5}$ can be calculated with excellent accuracy~\cite{RK_theo1,RK_theo3}, and this makes it possible to search for new physics by a precise $R_K$ measurement. 
In order to compare the experimental $R_K$ value with the SM prediction, the IB process of the $K_{e2}$ decay ($K_{e2\gamma}^{\rm IB}$) has to be included in the experimental $K_{e2}$ sample ($K_{e2(\gamma)}=K_{e2}+K_{e2\gamma}^{\rm IB}$) because it is impossible to experimentally separate the IB process from the $K_{e2}$ decay, especially in the soft-photon limit.
In the $R_K$ experiment, the $K_{e2\gamma}^{\rm SD}$ \color{black} events  in which no photon is detected \color{black} cannot be discriminated from the observed $K_{e2(\gamma)}$ sample.
Therefore, an accurate $Br(K_{e2\gamma}^{\rm SD})$ determination is very important for the $K_{e2 \gamma}^{\rm SD}$ subtraction in the $R_K$ analysis. 

The KLOE collaboration has reported~\cite{KLOE2009} an experimental result, \color{black} $R_{\gamma}=(1.483 \pm 0.066_{\rm{stat}} \pm 0.013_{\rm{syst}})\times 10^{-5}$,  \color{black} for the branching ratio of $Br(K_{e2\gamma}^{\rm SD})$ relative to that of the $K_{\mu 2}$ decay in the partial phase space  where the charged particle momentum and photon energy are higher than 200~MeV/$c$ and 10~MeV, respectively. 
On the other hand, the J-PARC E36 collaboration recently reported \color{black} a result  \color{black} for the $K_{e2\gamma}^{\rm SD}$  branching ratio relative to the $K_{e2(\gamma)}$ decay as $Br(K_{e2\gamma}^{\rm SD}) / Br(K_{e2(\gamma)})= 1.12\pm0.07_{\rm stat} \pm 0.04_{\rm syst}$ using a CsI(Tl) calorimeter for the radiative photon measurement~\cite{E36_CsI}.
This value was converted using the SM prediction of $R_K$ as \color{black} $Br(K_{e2\gamma}^{\rm SD})/Br(K_{\mu 2(\gamma)})=Br(K_{e2\gamma}^{\rm SD})/Br(K_{e2(\gamma)})\times R_K$ and corrected for the phase space reduction of $0.667\pm0.003$,  \color{black} resulting in 
$R_{\gamma}=(1.85 \pm 0.11_{\rm{stat}} \pm 0.07_{\rm{syst}})\times 10^{-5}$, which is $\sim$25\% higher than the KLOE result.
The $R_{\gamma}$ value will affect the presently most precise $R_K$ result reported by NA62~\cite{NA62-2013}, since the $K_{e2 \gamma}^{\rm SD}$ branching ratio obtained by KLOE  was used in the NA62 analysis.
The ChPT prediction at order $O(p^4)$, $R_{\gamma}=1.447 \times 10^{-5}$, is in good agreement with the KLOE result~\cite{biji1,biji2,KLOE2009}.
On the other hand, the E36 result~\cite{E36_CsI} is in agreement with a recent lattice QCD calculation, $R_{\gamma}=(1.74\pm 0.21)\times 10^{-5}$~\cite{lattice}. 
The form factors of the $K_{e2 \gamma}^{\rm SD}$ decay have been calculated using ChPT at order $O(p^6)$~\cite{chiPT-2008} and a gauged nonlocal chiral quark model (NL$\chi$QM)~\cite{hosaka1, hosaka2, hosaka3}, and these can also provide a $R_{\gamma}$ prediction.
In order to resolve the above experimental and theoretical situation, an additional determination of $Br(K_{e2 \gamma}^{\rm SD})/Br(K_{e2(\gamma)})$ was pursued by the E36 collaboration using sandwich detectors consisting of an alternating stack of plastic scintillators and lead plates referred to as gap sandwich counters (GSC) \color{black} for the photon measurement from the $K_{e2 \gamma}^{\rm SD}$ and $K_{e2(\gamma)}$ decays as an alternative \color{black} to the \color{black} CsI(Tl) detector used in~\cite{E36_CsI}.  
A new analysis was carried out accounting for the IB process not only in $K_{e2(\gamma)}$ but also now in the SD process \color{black} $K_{e2\gamma(\gamma)}^{\rm SD}$=$K_{e2\gamma}^{\rm SD}+(K_{e2\gamma}^{\rm SD})_{\gamma}^{\rm IB}$~\cite{eric}. \color{black}
Finally, the results of the GSC and the previous CsI(Tl) analyses \color{black} were \color{black} combined after revising the IB correction scheme for the latter to produce the final SD result of the E36 experiment.

Section~\ref{sec:sec2} presents the E36 setup including the GSC.
Section~\ref{sec:sec3} describes details of the analysis such as GSC efficiency calibration (\ref{sec:sec3-1}), SD event selection (\ref{sec:sec3-2}), detector acceptance (\ref{sec:sec3-3}), and $Br$ determination (\ref{sec:sec3-4}).
Section~\ref{sec:sec4} includes a discussion of systematic uncertainties, and the final results of the E36 experiment are discussed in Sec~\ref{sec:sec5}.


%
%
 \begin{figure}
  \includegraphics[width=.99\textwidth]
  {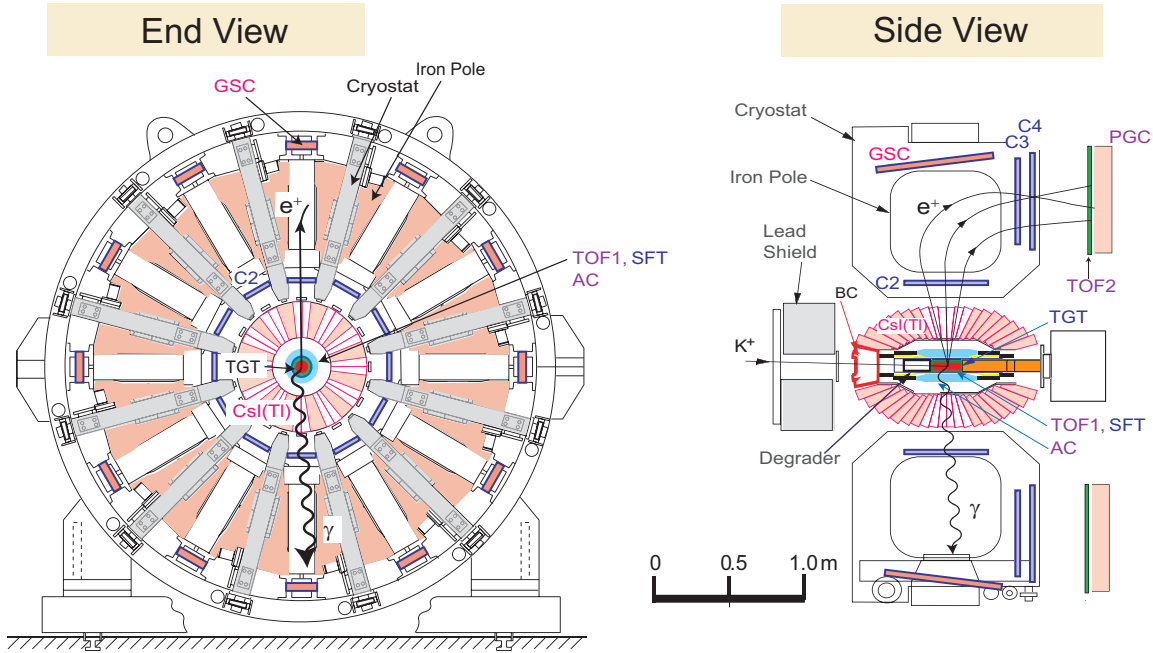}
  \caption{Schematic cross sectional side view (right) and end view (left) of the E36 detector configuration~\cite{E36_CsI}, which was originally constructed for the KEK-PS E246/470 experiments~\cite{e246, E246-2003, spectros1,spectros2,spectros3,spectros4,spectros5}. 
  The radiative photons from $K_{e2\gamma(\gamma)}^{\rm SD}$ were detected by GSC.
 \color{black} Charged particles were tracked and momentum-analyzed by reconstructing the particle trajectories using multi-wire proportional chambers located at the entrance and exit of the magnet gap in the toroidal spectrometer, as well as the segmented $K^+$ stopping target and a spiral fiber tracker. \color{black}
  Particle identification to accept $e^+$s and to reject $\mu^+$s was carried out using aerogel (AC), lead glass counters (PGC), and by measuring the time-of-flight between the TOF1 and TOF2 scintillator counters. 
  The SD events were observed with high-energy photon emission mainly in the opposite direction to the $e^+$ motion.}  
  \label{figset}
 \end{figure}

\section{The J-PARC E36 experiment} \label{sec:sec2}
\begin{figure}[hbtp]
 \includegraphics[width=.80\textwidth]{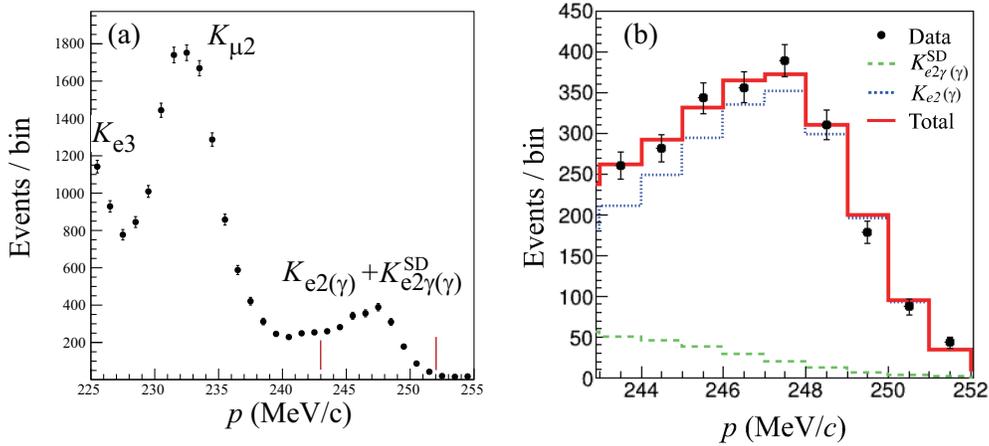}
 \caption{   
 Charged particle momentum spectrum with the $e^+$ selection condition \color{black} (a) in the region of 225$-$255 MeV/$c$ and (b) in the expanded region around the $K_{e2(\gamma)}$ peak. \color{black}
 The dots (black) are the experimental data. The dashed (green) and dotted (blue) lines are the decomposed $K_{e2\gamma(\gamma)}^{\rm SD}$ and $K_{e2(\gamma)}$ decays, respectively, with the shapes determined by simulations. 
 The solid (red) line is the fitted result obtained by adding both decay modes.
 \color{black} The tails of the $K_{e3}$ and $K_{\mu2}$ events do not extend beyond 243 MeV/$c$. \color{black} }
 \label{fig_Ke2.mom}
\end{figure}

\color{black} The E36 experiment~\cite{E36_CsI,E36SFT-2015-1,E36SFT-2015-2,E36AC-2015,E36PGC-2015} adopted a stopped $K^+$ method in conjunction with a 12-sector iron-core superconducting toroidal spectrometer for the charged particle measurement and a highly segmented CsI(Tl) calorimeter and GSCs for the photon measurement, as shown in Fig.~\ref{figset}.
Charged particles were tracked and momentum-analyzed by reconstructing the particle trajectories using multi-wire proportional chambers located at the entrance and exit of the magnet gap, as well as the segmented $K^+$ stopping target and a spiral fiber tracker. 
The momentum resolution was obtained to be $\sigma_p=2.0$ MeV/$c$ at 236 MeV/$c$.
Particle identification to accept $e^+$s and to reject $\mu^+$s was carried out using aerogel (AC), lead glass counters (PGC), and by measuring the time-of-flight between the TOF1 and TOF2 scintillator counters. 
It should be noted that the charged particle analysis was common for the $K_{e2(\gamma)}$ and $K_{e2\gamma(\gamma)}^{\rm SD}$ decays, while the photon detection was only required for the $K_{e2\gamma(\gamma)}^{\rm SD}$ selection.  \color{black} 
Details of the experimental methods, detector configuration, and analysis procedure for \color{black} charged particle momentum $(p)$ determination and particle identification (PID) are \color{black} described in Ref.~\cite{E36-CsI-2018}. 
\color{black} The dots in Fig.~\ref{fig_Ke2.mom} show the $p$ spectrum with the $e^+$ selection condition (a) in the region of 225$-$255 MeV/$c$ and (b) in the expanded region around the $K_{e2(\gamma)}$ peak indicated by the vertical lines in (a). 
The $K_{e2(\gamma)}$, $K_{e2\gamma(\gamma)}^{\rm SD}$, and $K_{e3}$ decays, as well as the remaining $K_{\mu2}$ events due to $\mu^+$ mis-identification can be seen \color{black} in Fig.~\ref{fig_Ke2.mom}(a).
\color{black} Since the $e^+$ selection conditions were relaxed to increase the $K_{e2\gamma(\gamma)}^{\rm SD}$  statistics \color{black} taking advantage of the good timing resolution of the GSC (see below), the $K_{\mu2}$ yield was about 2.3 times higher than the result obtained in the CsI(Tl) analysis~\cite{E36_CsI}. \color{black}
\color{black} 
The $e^+$ events selected without the GSC requirement were used to obtain the sample of the $K_{e2(\gamma)}+K_{e2\gamma(\gamma)}^{\rm SD}$ events.
The subset of these events selected by applying the GSC requirement provided a sample enriched in $K_{e2\gamma(\gamma)}^{\rm SD}$, and the momentum distributions for these event samples were fit simultaneously to the simulated distribution with the total number of $K_{e2(\gamma)}$ and $K_{e2\gamma(\gamma)}^{\rm SD}$ as free parameters \color{black} (see Sec~\ref{sec:sec3-4}). \color{black}
The dashed (green) and dotted (blue) lines in Fig.~\ref{fig_Ke2.mom}(b) are the decomposed $K_{e2\gamma(\gamma)}^{\rm SD}$ and $K_{e2(\gamma)}$ decays determined by the fitting, and the solid (red) line is the fitted result obtained by adding both decay modes. 
\color{black}

The CsI(Tl) calorimeter, an assembly of 768 CsI(Tl) crystals, covered 70\% of the total solid angle. 
There were 12 holes for the outgoing charged particles to enter the spectrometer and 2 holes for the beam entrance and exit, thus not all of the radiated photons from $K^+$ decays can be detected by the calorimeter.
However, some photons passed through the holes of the CsI(Tl) calorimeter, and a fraction of these escaping photons entered \color{black} the GSCs,   \color{black} which were located at the outer radius of the magnet pole in each sector, as shown in Fig.~\ref{figset}.
The \color{black} GSCs \color{black} were constructed with four layers of Pb and plastic scintillator. \color{black}
The size of each layer was 900~mm$\times$196~mm and the thickness was 3.7~mm (10~mm) for Pb (plastic) corresponding to 2.7 radiation lengths. 
The solid angle of each CsI(Tl) hole was 15$^{\circ}$ in the azimuthal direction and 45$^{\circ}$ in the polar direction, while the GSC covered only 6$^{\circ}$ (azimuthal) and 32$^{\circ}$ (polar), \color{black} and about 30\% of the photons escaping through the CsI(Tl) holes reached the GSCs.  \color{black}
The electromagnetic showers generated by the interaction of the emitted photons with the materials of the magnet poles or CsI(Tl) calorimeter could also produce signals in the GSC.
The scintillation light from the four plastic layers was transported to a photomultiplier tube through an acrylic light guide in each unit. 
Due to spatial restrictions, the scintillation photons were collected with a photomultiplier only on one end (upstream) of the detector. 

Since photons from the $K_{e2 \gamma(\gamma)}^{\rm SD}$ decays can be detected either by the GSC or the CsI(Tl) calorimeter, the $Br(K_{e2 \gamma(\gamma)}^{\rm SD})/Br(K_{e2(\gamma)})$ ratio from the GSC measurement can be compared with the result of the CsI(Tl) analysis~\cite{E36_CsI}.
In the current work, the data sets from the same running period were used as in~\cite{E36_CsI}.
It should be emphasized that information of the photon energy and hit position were used in the CsI(Tl) analysis for reconstructing the $K_{e2\gamma(\gamma)}^{\rm SD}$ kinematics in the event selection~\cite{E36_CsI}. 
On the other hand, although \color{black} the GSC  \color{black} could not provide the energy and hit position of the radiative photons, the $e^+$ momentum spectrum of the $K_{e2 \gamma(\gamma)}^{\rm SD}$ decays was obtained by requiring a GSC hit without imposing any kinematical constraints.
In addition, since the intrinsic timing resolution of the GSC was $\sim$1~ns, which was about 1/10 that of the CsI(Tl), and the GSC singles rate was much lower than the CsI(Tl) singles rate due to the smaller solid angle, the $K_{\mu2}$ \color{black} background events  \color{black} with an accidental photon detector hit present in the CsI(Tl) based analysis was strongly suppressed in the GSC analysis. 
As a result, the PID condition for the $\mu^+$ rejection in the GSC analysis could be relaxed, and the $e^+$ PID detection efficiency was nearly 100\%, while it was only $\sim$$75\%$ in the CsI(Tl) analysis~\cite{E36_CsI}.
The detection efficiency of the GSC for radiative photons is less than 100\%, due to the limited probability of the photon interaction with the lead materials and the hardware threshold for the signal readout. 
The GSC efficiency is defined as the product of the individual efficiencies due to these two effects.
The former effect was taken into account in the Monte Carlo (MC) simulation using the known photon cross section, while the latter effect can be corrected for in the analysis (see below and Eq.~\ref{brform}).

\section{Analysis}\label{sec:sec3}
\subsection{GSC efficiency}\label{sec:sec3-1}
\begin{figure*}[htp]
\center
\includegraphics[width=.50\textwidth]{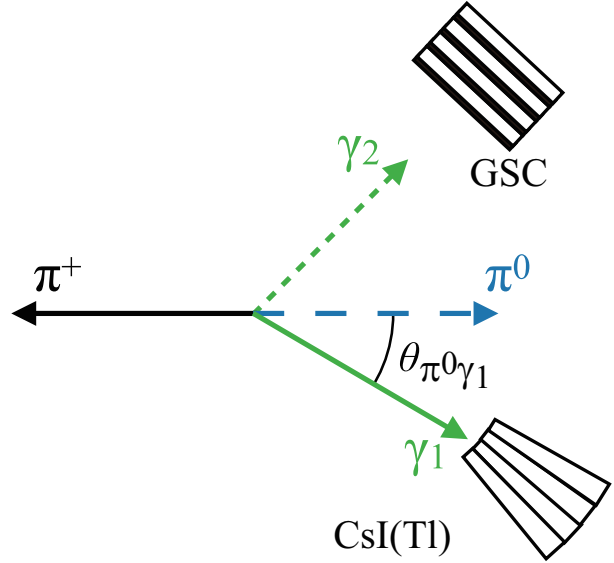}
\caption{
The $K_{\pi2}$ events used for the determination of the GSC efficiency reduction.  
In order to reduce effects of shower leakage, the $\gamma_1$ energy was calculated from the $\gamma_1$ and $\pi^+$ directions assuming the $K_{\pi2}$ decay kinematics \color{black} rather than using cluster energies deposited in the hit modules, which was  substantially affected by shower leakage. \color{black}
Then, the $\gamma_2$ photon energy and direction were calculated from the $\gamma_1$ and $\pi^+$ information.}
\label{Kpi2_kine}
\end{figure*}

\begin{figure*}[htp]
\center
\includegraphics[width=.80\textwidth]{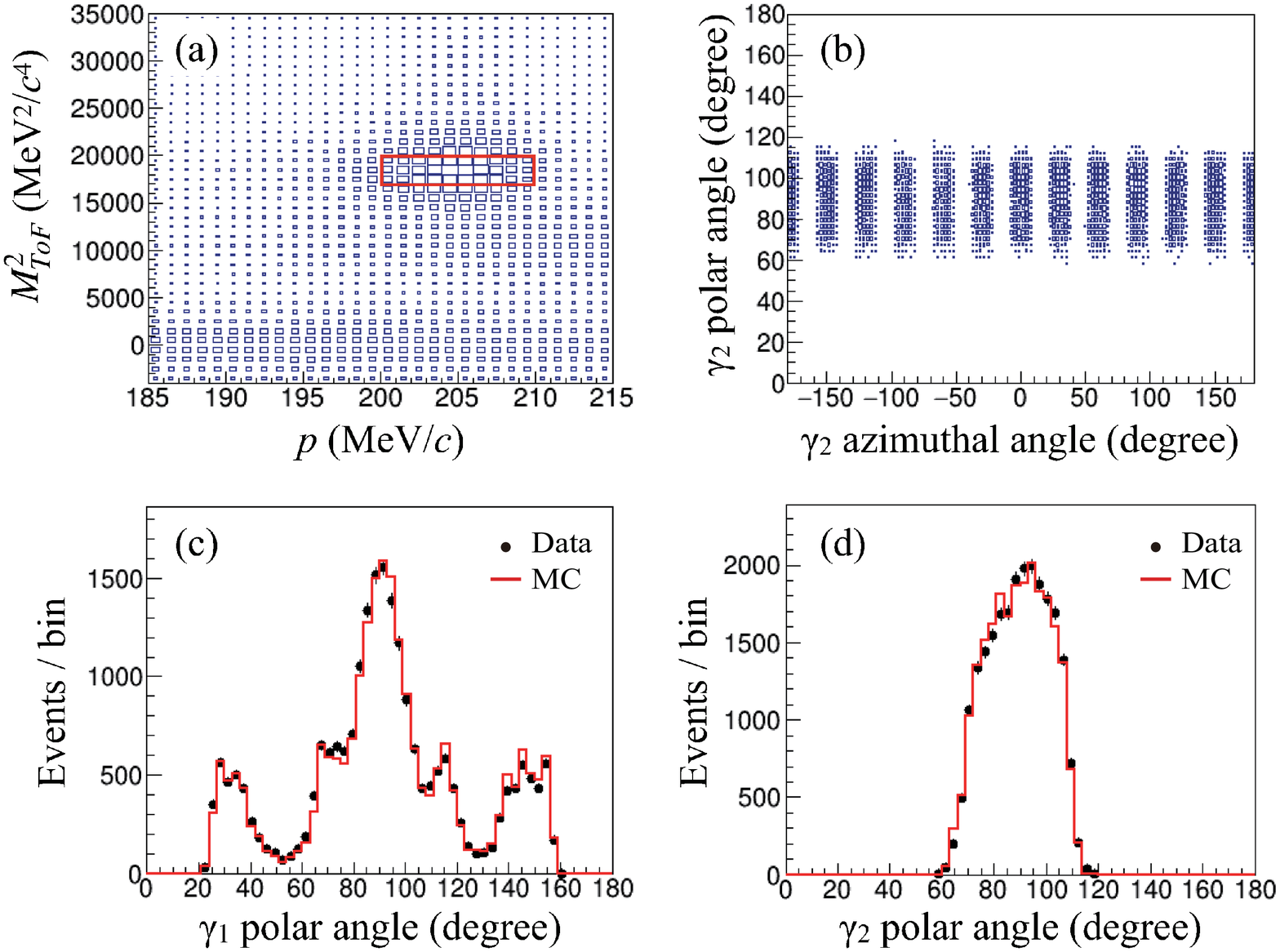}
\caption{
The $K_{\pi 2}$ decays were selected by the cut window of the $(p, M^2_{\rm TOF})$ correlation, as shown in (a).
Then, requiring the 1-cluster hit in the CsI(Tl) calorimeter, the $K_{\pi2}$ events  were further selected: (b) polar-azimuthal angular distribution of the $\gamma_2$ photon, and polar angle distributions of  the (c) $\gamma_1$ and (d) $\gamma_2$ photons.
The simulation histograms (red line) in (c) and (d) are commonly normalized by requiring the total yield of each histogram is the same as that of the experimental data (dots). 
}
\label{fig_gamma2}
\end{figure*}

The GSC efficiency reduction due to the hardware threshold was determined using \mbox{$K^+ \rightarrow \pi^+ \pi^0$ ($K_{\pi2}$)}, as shown in Fig.~\ref{Kpi2_kine}, by comparing the experimental data with the MC simulation in which \color{black} no threshold for the energy deposit was simulated. \color{black} 
This determination of the \color{black} efficiency reduction was used to calculate the $Br(K_{e2\gamma(\gamma)}^{\rm SD})/Br(K_{e2(\gamma)})$ value.  \color{black}
The stopped $K_{\pi2}$ decays with the $\pi^+$ tracked, one photon detected by the CsI(Tl) barrel, and one escaping photon passing through the holes were selected.
The charged particle analysis was used to determine the $\pi^+$ momentum and direction with the conditions of $200<p<210$~MeV/$c$ and $130^2<M^2_{\rm TOF}<140^2$~${\rm MeV}^2/c^4$, as shown in Fig.~\ref{fig_gamma2}(a), where $M^2_{\rm TOF}$ is the mass-squared of the charged particle obtained from time-of-flight, momentum, and path length.
\color{black} Only \color{black} events with one photon cluster ($\gamma_1$) detected by the CsI(Tl) calorimeter were chosen,  and the information of the escaping photon ($\gamma_2$) was calculated by imposing  the $K_{\pi 2}$ kinematics, as shown in Fig.~\ref{Kpi2_kine}.
\color{black} An electromagnetic shower generated in the CsI(Tl) calorimeter cannot be 100\% contained and a part of photon energy was lost due to $\gamma$'s escaping from the calorimeter as well as the threshold effect in each module~\cite{E36_CsI}. 
\color{black} Due to this shower leakage effect, the $\pi^0$ invariant mass calculated from the observed photon energy was shifted to 95 MeV/$c^2$ from 135 MeV/$c^2$. \color{black}
In order to reduce the effects of this shower leakage in the energy determination, the $\gamma_1$ energy ($E_{\gamma_1}$) was calculated from the opening angle between the $\gamma_1$ and $\pi^0$ directions ($\theta_{\pi^0 \gamma_1}$) assuming $K_{\pi2}$ kinematics, and the corresponding $\gamma_2$ energy ($E_{\gamma_2}$) was computed as
\begin{subequations}
\begin{align}
E_{\gamma_{1}}&=M_{\pi^0}^2/(2p_{\pi^0})\frac{1}{\sqrt{1+(M_{\pi^0}/p_{\pi^0})^2}-{\rm cos}\theta_{\pi^0 \gamma_{1}}} \\
E_{\gamma_{2}}&=E_{\pi^0}-E_{\gamma_{1}},
\end{align}
\end{subequations}
where $E_{\pi^0}$=245~MeV and $p_{\pi^0}$=205~MeV/$c$ are the monochromatic $K_{\pi2}$ $\pi^0$ relativistic energy and momentum, respectively, and $M_{\pi^0}$=135~MeV/$c^2$ is the $\pi^0$ rest mass.
\color{black} Here it should be noted that the accidental backgrounds were taken into account in the simulation~\cite{E36_CsI} to reproduce the experimental $K_{\pi2 }$ distributions.
Since the $K_{\mu2}$ decays did not have accompanying photons, the CsI(Tl) signals which coincided with the $K_{\mu2}$ decays can be treated as pure accidental backgrounds, and these events were merged with the simulation data.
The pile-up probability for the accidental backgrounds was obtained to be 18.85$\pm$0.03\%, resulting in 2.24\% \color{black} of $K_{\pi2}$ contamination in the MC sample.
The validity of this simulation method was carefully checked using two photons from the $K_{\pi2}$ decay, and the simulation calculations were in good agreement with the experimental data. \color{black}
For this check, \color{black} the $\gamma_2$ photons were further selected by requiring the photon passage through the CsI(Tl) holes using the $\gamma_2$ direction and the $K^+$ decay vertex, as shown in Fig.~\ref{fig_gamma2}(b).
The dots shown in Fig.~\ref{fig_gamma2}(c) and (d) are the $\gamma_1$ and $\gamma_2$ polar angle distributions, respectively, and the red lines are the simulation calculations.
The data and simulation are in good agreement, which indicates that the $\gamma_2$ photon kinematics were correctly determined from the $\gamma_1$ and $\pi^+$ information.
The energy and polar \color{black} angular resolution of the $\gamma_2$ photon was estimated from the simulation to be 4 MeV and 3.6$^{\circ}$, \color{black} respectively.

\begin{figure*}[htp]
\center
\includegraphics[width=.80\textwidth]{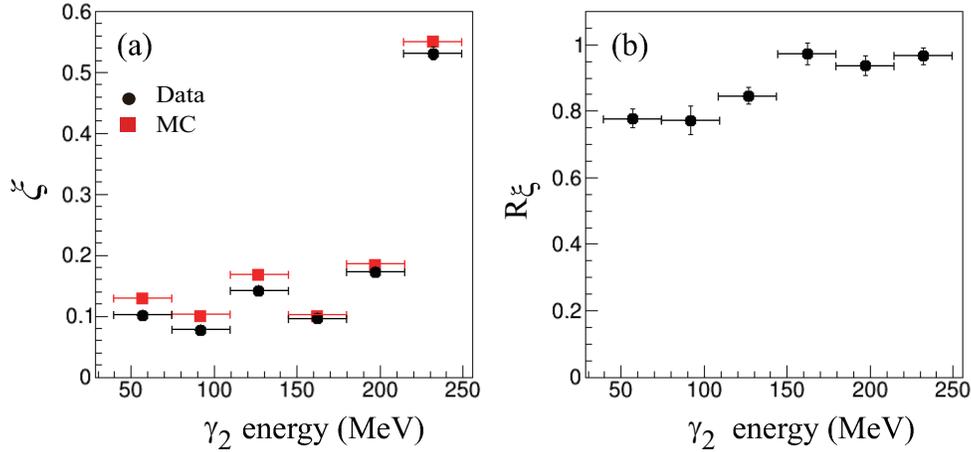}
\caption{
(a) The $\xi(E_{\gamma_2})$ values obtained as a function of the photon energy for the experiment (circle/black) and simulation (square/red) data, and (b) the ratio of the experimental and simulated $\xi$ values, $R_\xi(E_{\gamma_2})=\xi^{\rm Exp}(E_{\gamma_2})/\xi^{\rm MC}(E_{\gamma_2})$. \color{black} The data points \color{black} are obtained using the $K_{\pi 2}$ events with one photon escaping through the CsI(Tl) holes. 
 }  
\label{fg_gv_eff}
\end{figure*}

A quantity $\xi$, which is the product of the GSC acceptance relative to the solid angle of the CsI(Tl) hole and the GSC detection efficiency, was defined as
\begin{eqnarray}
\xi(E_{\gamma_2})=N^{\rm GSC}(E_{\gamma_2})/N^{\rm EP}(E_{\gamma_2}), \label{eq_xi}
\end{eqnarray}
where $N^{\rm EP}$ and $N^{\rm GSC}$ are the numbers of events with a photon passing through the holes obtained in the above analysis and the actual GSC hit events within a coincidence window $\pm$10 ns from the $K^+$ decay, respectively. 
\color{black} Since the solid angle of the GSC was smaller than that of the CsI(Tl) hole, some photons hit the CsI(Tl) module around the hole and magnet pole without directly hitting the GSC. 
In addition, these photons can generate showers from the CsI(Tl) grazing and magnet pole scattering and create signals in the GSC, which were also included in the $N^{\rm GSC}$ counts. \color{black}
Next, the $\xi$ ratio, 
\begin{eqnarray}
R_\xi(E_{\gamma_2})=\xi^{\rm Exp}(E_{\gamma_2})/\xi^{\rm MC}(E_{\gamma_2}),
\end{eqnarray}
corresponding to the efficiency reduction due to the hardware threshold for the GSC signals, was determined as a function of $E_{\gamma 2}$. 
Note that effects due to accidental backgrounds are suppressed to leading orders in this ratio and estimated to be negligibly small. \color{black}
Figures~\ref{fg_gv_eff}~(a) and (b) show the $\xi(E_{\gamma_2})$ and $R_\xi(E_{\gamma_2})$ distributions, respectively.
The trend of the experimental $\xi(E_{\gamma_2})$ distribution is in agreement with that of the simulation calculation.
The $\xi$ values below 200 MeV are in the range of 10$-$20\%, because the GSC coverage for the escaping photons is $\sim$30\% and the photon conversion probability is $\sim$70\%.
Also, the contribution of \color{black} non-observed  \color{black} photon clusters in the CsI(Tl) due to the inefficiency reduced the $\xi$ value, since some of the two-cluster events were mistakenly identified as one-cluster events and contributed to the denominator of Eq.~\ref{eq_xi}, $N^{\rm EP}$.
\color{black} This fraction was estimated to be \color{black} $\sim$30\% \color{black} of the total events according to the MC simulation.  \color{black}
On the other hand, the opening angle between the $\pi^+$ and the $\gamma_2$ photon with an energy higher than 200 MeV is strongly peaked in the back-to-back direction because of the $K_{\pi2}$ kinematics.
Hence, most of the $\gamma_2$ photons were directed toward the GSC and the $\xi$ value is $\sim$50\%.
Since the photon conversion probability was taken into account in the simulation, the $R_\xi(E_{\gamma_2})$ distribution has a nearly flat structure over the entire photon energy region.
\color{black} The $R_\xi$ drop from unity in the low energy region was interpreted as the efficiency reduction due to the hardware threshold effect for the GSC signal readout. \color{black}
An uncertainty of the photon interaction cross section with the GSC material also affects the $\xi$ and $R_\xi$ determination, \color{black} but this effect is only at the $\sim$1\% level and it is much smaller than the above threshold effect. \color{black}
In the $R_\xi$ correction in the $K_{e2\gamma(\gamma)}^{\rm SD}$ analysis, a constant value of the data points within the interval of each bin was assumed, as indicated by the horizontal bars in Fig.~\ref{fg_gv_eff}.

\subsection{${K_{e2\gamma(\gamma)}^{\rm SD}}$ event selection}\label{sec:sec3-2}
\begin{figure}[hbtp]
 \includegraphics[width=.80\textwidth]{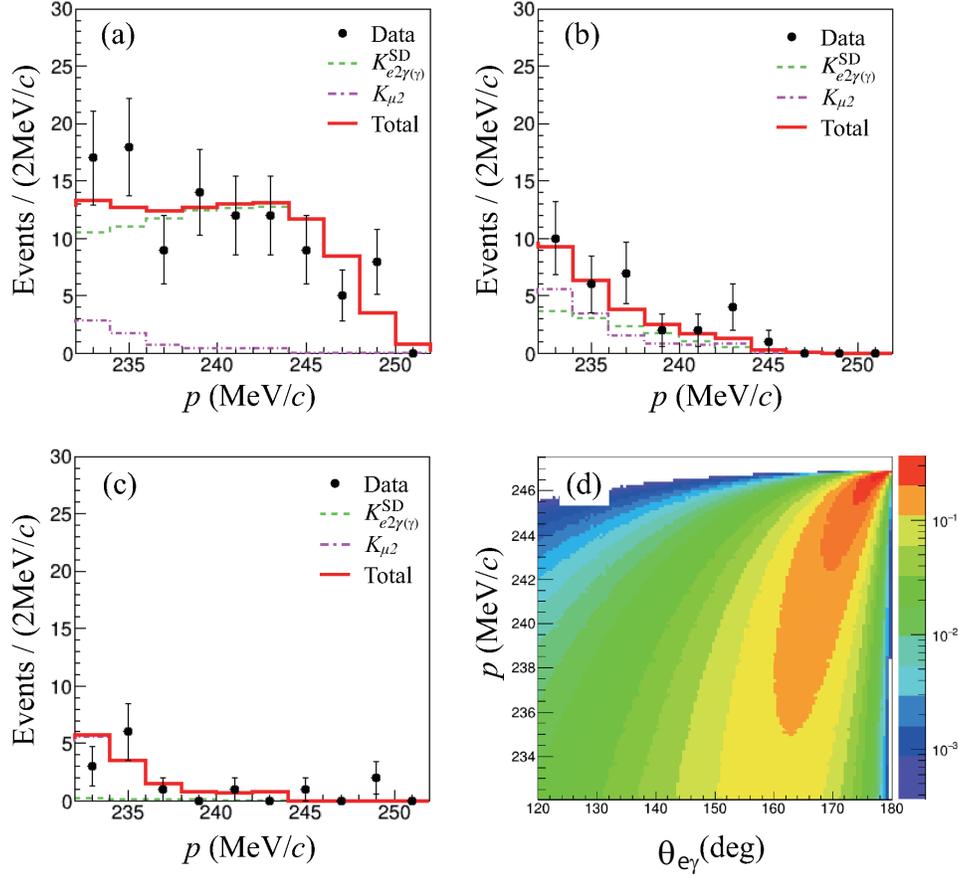}
 \caption{   
 Charged particle momentum spectra  for the (a) $\Delta$sec=6, (b) $\Delta$sec=$\pm$5, and (c) $\Delta$sec=$\pm$4 requirements.
 The dots (black) are the experimental data. 
 The dashed (green) and dashed-dotted (magenta) lines are the $K_{e2\gamma(\gamma)}^{\rm SD}$ and $K_{\mu2}$ decays, respectively, determined by simulation calculations.
 \color{black} The azimuthal range of events selected with $\Delta$sec=6 is $180^{\circ}\pm15^{\circ}$. \color{black}
 The solid (red) line is the fitted result obtained by adding the two decay modes. 
 Plot (d) shows the theoretical correlation for the $e^+$ momentum and the ($e^+$,$\gamma$) opening angle for $K_{e2\gamma(\gamma)}^{\rm SD}$ decays. \color{black}
 \color{black} Due to the finite momentum resolution of the $e^+$ measurement, the data points extend beyond the kinematic $K_{e2\gamma(\gamma)}^{\rm SD}$ endpoint of 247 MeV/$c$.  \color{black}
 }
 \label{fig.mom}
\end{figure}

By selecting $e^+$ in the charged particle tracking and PID~\cite{E36_CsI}, as well as requiring the GSC to fire \color{black} and the CsI(Tl) not to fire,  \color{black} the $K_{e2 \gamma(\gamma)}^{\rm SD}$ events \color{black} with a small contribution from the $K_{\mu 2}$ decay with accidental background \color{black} were successfully observed mostly in opposite sectors (i.e. back-to-back), while the $K_{e2(\gamma)}$ events were \color{black} rejected by the GSC requirement \color{black} and negligible, as shown in Fig.~\ref{fig.mom}.
\color{black} The $K_{\mu2 \gamma}$ decay with a radiative photon emission was also negligible due to the $\mu^+$ suppression by the PID systems. \color{black}
The $e^+$ momentum spectra were obtained by selecting events with a difference in the spectrometer sector number ($\Delta$sec) between the accepted $e^+$ and \color{black} the hit in the GSC.  \color{black}
Two neighboring sectors differ by 30$^{\circ}$ in azimuthal angle.
In Fig.~\ref{fig.mom}, the spectra for (a) $\Delta$sec=6, (b) $\Delta$sec=$\pm$5, and (c) $\Delta$sec=$\pm$4 are shown corresponding to differences in \color{black} azimuthal angle of \color{black} $\sim$180$^{\circ}$, $\sim$150$^{\circ}$, and $\sim$120$^{\circ}$.
A schematic view of a typical $\Delta$sec=6 event is drawn in  Fig.~\ref{figset}.
The $K_{e2 \gamma(\gamma)}^{\rm SD}$ events are observed only for $\Delta$sec=6, $\pm$5 and negligibly for $\Delta$sec=$\pm$4, as seen in Fig~\ref{fig.mom}, \color{black} indicating that they are concentrated in the back-to-back direction of the $e^+$ and photon momenta.
On the other hand, the peak structure due to remaining $K_{\mu2}$ backgrounds is seen in the spectra for $\Delta$sec=$\pm$5 and $\Delta$sec=$\pm$4 \color{black} in Figs.~\ref{fig.mom}(b) and (c), respectively. 
Since the $e^+$ endpoint momentum of the $K_{e3}$ decay is 228 MeV/$c$, the $K_{e3}$ contribution with one photon hit in the GSC is negligible taking into account the $e^+$ momentum resolution.  \color{black}
Figure~\ref{fig.mom}(d) shows the theoretical correlation plot for the $e^+$ momentum and the ($e^+$,$\gamma$) opening angle for $K_{e2\gamma(\gamma)}^{\rm SD}$ decays, \color{black} which is consistent with the observed $K_{e2\gamma(\gamma)}^{\rm SD}$ yields shown in Fig.~\ref{fig.mom}(a)(b)(c).
It should be noted the $e^+$ momentum range of events with $\Delta$sec=6 is higher than that with $\Delta$sec=$\pm$5, which is also consistent with Fig.~\ref{fig.mom}(d), validating the treatment  of the SD dynamics in the \color{black} simulation. \color{black}
\color{black} Since bremsstrahlung photons generated from the $e^+$ interaction with the target materials and the IB photons from the $K_{e2\gamma(\gamma)}^{\rm SD}$ and $K_{e2(\gamma)}$ decays are nearly parallel to the $e^+$ motion and they could be accepted by the GSC, a significant contribution to events with $\Delta$sec=0 was observed, but this did not disturb the $K_{e2\gamma(\gamma)}^{\rm SD}$ measurement.  \color{black}

\subsection{Detector acceptance}\label{sec:sec3-3}
The detector acceptances for the $K_{e2\gamma(\gamma)}^{\rm SD}$, $K_{e2(\gamma)}$, and $K_{\mu2}$ decays were calculated by a Geant4-based Monte Carlo simulation assuming the theoretical scheme of vector and axial-vector transitions~\cite{biji1,biji2,rev_mod}. 
Here, the vector form factor, $V$, was assumed to have a momentum  transfer dependence $V=V_0[1+\lambda(1-2E_{\gamma}/m_K)]$, according to ChPT at order $O(p^6)$~\cite{chiPT-2004,chiPT-2008}, where $E_{\gamma}$ is the photon energy, $m_K=$ is the kaon mass, and $\lambda$ is the slope parameter which was taken to be $\lambda=0.3\pm0.1$~\cite{rev_mod}.
In order to avoid the infra-red divergence problem\color{black}~\cite{weinberg}, \color{black} the effect of the IB process was calculated following~\cite{gatti} for $K_{e2(\gamma)}$ and~\cite{eric} for $K_{e2\gamma(\gamma)}^{\rm SD}$ using a re-summation scheme of multiple photon emission.
Since these IB photons were not observed using the photon detectors in the present work, \color{black} the $e^+$ acceptance loss due to the IB emission was taken into account for the $K_{e2(\gamma)}$ and $K_{e2\gamma(\gamma)}^{\rm SD}$ acceptance determination. 
Details of the simulation calculation are described in~\cite{E36_CsI}.

\subsection{$Br(K_{e2\gamma(\gamma)}^{\rm SD})/Br(K_{e2(\gamma)})$ determination}\label{sec:sec3-4}
In the present study, the $K_{e2 \gamma(\gamma)}^{\rm SD}$ branching ratio relative to the $K_{e2(\gamma)}$ decay was obtained from the $K_{e2\gamma(\gamma)}^{\rm SD}$ and $K_{e2(\gamma)}$ yields corrected for the detector acceptance and the GSC efficiency $R_\xi$ as 
\begin{eqnarray}
\frac{Br(K_{e2\gamma(\gamma)}^{\rm SD})}{Br(K_{e2(\gamma)})}
= \frac{N(K_{e2\gamma(\gamma)}^{\rm SD}) }{ N(K_{e2(\gamma)})} \cdot R_\Omega  \cdot \frac{1}{\langle R_\xi \rangle} 
, \label{brform}
\end{eqnarray}
where $N$ is the number of the accepted events of each decay mode, and $R_\Omega=\Omega(K_{e2(\gamma)})/\Omega(K_{e2\gamma(\gamma)}^{\rm SD})$ is the ratio of the overall acceptances $(\Omega)$ calculated by the MC simulation and obtained to be $R_{\Omega}=20.6\pm0.3$. \color{black} 
The $R_{\Omega}$ uncertainty is dominated by the systematics due to imperfect reproducibility of the detector acceptance in the simulation (see Sec~\ref{sec:sec4} ). \color{black}
The quantity $\langle R_\xi \rangle=0.92\pm 0.04$ is the average $R_\xi$ value weighted by the \color{black} theoretical  \color{black} SD photon energy spectrum using the MC data.
\color{black} It is to be noted that the IB process is included in both $K_{e2(\gamma)}$ and $K_{e2\gamma(\gamma)}^{\rm SD}$ samples described in Eq.~\ref{brform}.  \color{black}
The spectrum in Fig.~\ref{fig_Ke2.mom}(b) was decomposed by simulating the shapes of the $K_{e2\gamma(\gamma)}^{\rm SD}$ and $K_{e2(\gamma)}$ spectra and fitting their linear combination to the measured spectrum.
To further constrain $Br(K_{e2\gamma(\gamma)}^{\rm SD})$, \color{black} the $K_{e2\gamma(\gamma)}^{\rm SD}$ events shown in Fig.~\ref{fig.mom} were used.
The Fig.~\ref{fig.mom}(a)$-$(c) and Fig.~\ref{fig_Ke2.mom}(b) spectra obtained with the GSC hit requirement and without the GSC constraint, respectively, \color{black} were fit simultaneously with the ratio $Br(K_{e2\gamma(\gamma)}^{\rm SD})/Br(K_{e2(\gamma)})$ defined in Eq.~\ref{brform}, using the yield of $K_{e2\gamma(\gamma)}^{\rm SD}$ and $K_{\mu2}$ decays in the $\Delta$sec=6 data as free parameters.
\color{black} The relative normalization of the MC spectra in the $\Delta$sec=6, $\pm 5$, and $\pm 4$ distribution was fixed, based on the MC calculation. \color{black}
The $K_{\mu2}$ yield for events with $\Delta$sec=$\pm$4,$\pm$5 was \color{black} assumed \color{black} to be twice that for events with $\Delta$sec=6. 
The four momentum spectra of Fig.~\ref{fig_Ke2.mom}(b), Fig.~\ref{fig.mom}(a), (b), and (c) were simultaneously reproduced by this fitting. 
The fitted results are shown by the solid (red) lines, and the decomposed $K_{e2 \gamma(\gamma)}^{\rm SD}$, $K_{e2(\gamma)}$, and $K_{\mu2}$ decays are indicated by the dashed (green), dotted (blue), and dashed-dotted (magenta) lines,  respectively. 
The number of accepted $K_{e2\gamma(\gamma)}^{\rm SD}$ events was obtained to be $N(\Delta$sec=6)$=95\pm10$ and $N(\Delta$sec=$\pm$5)$=13\pm1$, and  $N(K_{e2(\gamma)})$ is $1939 \pm 49$.
The $Br(K_{e2\gamma(\gamma)}^{\rm SD})/Br(K_{e2(\gamma)})$ value was obtained to be $1.25\pm0.14$ with a reduced $\chi^2/$dof value $=66.0/58$.

\begin{figure}[hbtp]
 \includegraphics[width=.5\textwidth]{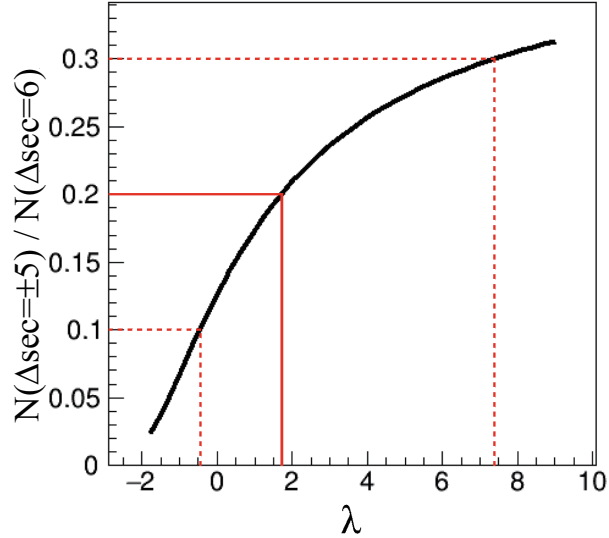}
 \caption{The $N(\Delta$sec=$\pm$5)$/N(\Delta$sec=6) value calculated using the simulation data as a function of the $\lambda$ parameter. 
 The $\lambda$ value is obtained from the experimental result of $N(\Delta$sec=$\pm$5)/$N(\Delta$sec=6) to be $1.7^{+5.7}_{-2.2}$.
 }
 \label{fig.g5g6}
\end{figure}

\color{black} In order to determine the $\lambda$ parameter, the $K_{e2\gamma(\gamma)}^{\rm SD}$ yields were once again obtained without the constraint on the relative normalization of the MC spectra in the $\Delta$sec=6, $\pm 5$ distribution as $N(\Delta$sec=6)$=96\pm11$ and $N(\Delta$sec=$\pm$5)$=20\pm9$, and
\color{black} the ratio $N(\Delta$sec=$\pm$5)/$N(\Delta$sec=6) was observed to be \color{black} 0.2$\pm0.1$. \color{black}
The $\lambda$ parameter can be determined from this ratio, because the ($e^+,\gamma $) angular distribution depends on $\lambda$. \color{black}
It should be noted that the detection efficiency of each GSC is cancelled out in this ratio by integrating over all the data obtained in the 12 sectors.
The $N(\Delta$sec=$\pm$5)/$N(\Delta$sec=6) value calculated using the simulation data as a function of the $\lambda$ parameter is shown in Fig.~\ref{fig.g5g6}.
\color{black} The $N(\Delta$sec=$\pm$5)/$N(\Delta$sec=6) value  increased by 5\% due to \color{black} small efficiency differences originating from a change in the photon energy distribution for the $\Delta$sec$\pm$5 and $\Delta$sec=6.  
Although the statistical uncertainty is large, the $\lambda$ value is obtained to be $1.7^{+5.7}_{-2.2}$, which is consistent with the theoretical calculation~\cite{rev_mod} and the input value in the MC simulation.

\begin{table}[htbp]
\center
\caption{Summary of the systematic uncertainties for the $Br(K_{e2 \gamma(\gamma)}^{\rm SD})/Br(K_{e2(\gamma)})$ ratio determination.
 Items except for the GSC detection efficiency, GSC timing window, and $K_{\mu 2}$ background subtraction contributed to the $R_{\Omega}$ uncertainty. \color{black} }
\begin{tabular}{l r}
\hline
Source & Uncertainty \\ \hline

GSC detection efficiency $\langle R_{\xi}\rangle$            & 0.060\\  
GSC misalignment 			        & $<0.001$\\
GSC timing window	            	& 0.025\\
$K_{\mu2}$ background subtraction       & 0.042\\
AC detection efficiency  				& 0.008\\ 
PGC detection efficiency  				& 0.010\\
TOF detection efficiency  				& 0.013\\
$K_{e2\gamma(\gamma)}^{\rm SD}$ form factor 	& 0.001\\
$K^+$ stopping distribution             & 0.009\\
Material thickness in the central parts   & $<0.001$\\
Positron momentum resolution 	        & 0.002\\
Magnetic field  	        & 0.002\\
In-flight kaon decay 				    & 0.002\\
\hline
Total & 0.080\\
\hline
\end{tabular}
\label{sys.error.summary}
\end{table}

\section{Systematic uncertainties}\label{sec:sec4}
The systematic uncertainties for the $Br(K_{e2\gamma(\gamma)}^{\rm SD})/Br(K_{e2(\gamma)})$ determination are summarized in Table~\ref{sys.error.summary}.
Since the photon detection was only required for the $K_{e2\gamma(\gamma)}^{\rm SD}$ selection, the dominant contribution to the systematic uncertainty is due to the ambiguity of the radiative photon measurement. \color{black}
The $R_\xi$ values were calculated by changing the $K_{\pi2}$ selection conditions, and the parameter shifts were treated as the systematic uncertainty of the $R_\xi$ determination.
Even if there was an incorrect input of the interaction cross section in the simulation, the effect was included in the $R_\xi$ measurement and it does not affect the final result.
On the other hand, the effect of a GSC position misalignment was evaluated  by considering the maximum conceivable shift of 2 mm.
Although the accidental backgrounds in the GSC were highly suppressed, these contributions were checked by changing the selection window of the timing gate \color{black} from 8 ns to 12 ns, resulting in a 50\% increase in the accidental events. \color{black}
The effect from the $K_{\mu 2}$ subtraction was estimated by intentionally changing the $K_{\mu2}$ fraction with various PID selection conditions.
\color{black} The $K_{\mu2}$ yields changed from 50\% to 200\% compared with the standard selection condition, and  \color{black} the $Br(K_{e2\gamma(\gamma)}^{\rm SD})/Br(K_{e2(\gamma)})$ changes were interpreted as the contribution from this effect.
The momentum dependence \color{black} of the efficiency of \color{black} the PID detectors from 200 to 250~MeV/$c$ was measured using the $K_{e3}$ and in-flight $K_{e3}$ events and taken into account in the simulation. 
However, its statistical uncertainty introduced a possible change in the efficiency correction, which was regarded as a systematic effect in the efficiency correction~\cite{E36_CsI}. 
The ($e^+$, $\gamma$) angular correlation depends significantly on the $\lambda$ parameter, which introduces a systematic uncertainty through a change in the detector acceptances. 
The $Br(K_{e2\gamma(\gamma)}^{\rm SD})/Br(K_{e2(\gamma)})$ shift due to a parameter change of the theoretical uncertainty~\cite{rev_mod} $\Delta \lambda=0.1$ was interpreted as the systematic uncertainty.
In addition, effects from the $K^+$ stopping distribution, material thickness in the central parts of the detector, $e^+$ momentum resolution, magnetic field, and in-flight $K^+$ decay were also considered.
The total size of the systematic uncertainty in the $Br(K_{e2\gamma(\gamma)}^{\rm SD})/Br(K_{e2(\gamma)})$ determination was obtained by adding each item in quadrature to be 0.08.

\begin{figure}[hbtp]
 \includegraphics[width=1.\textwidth]{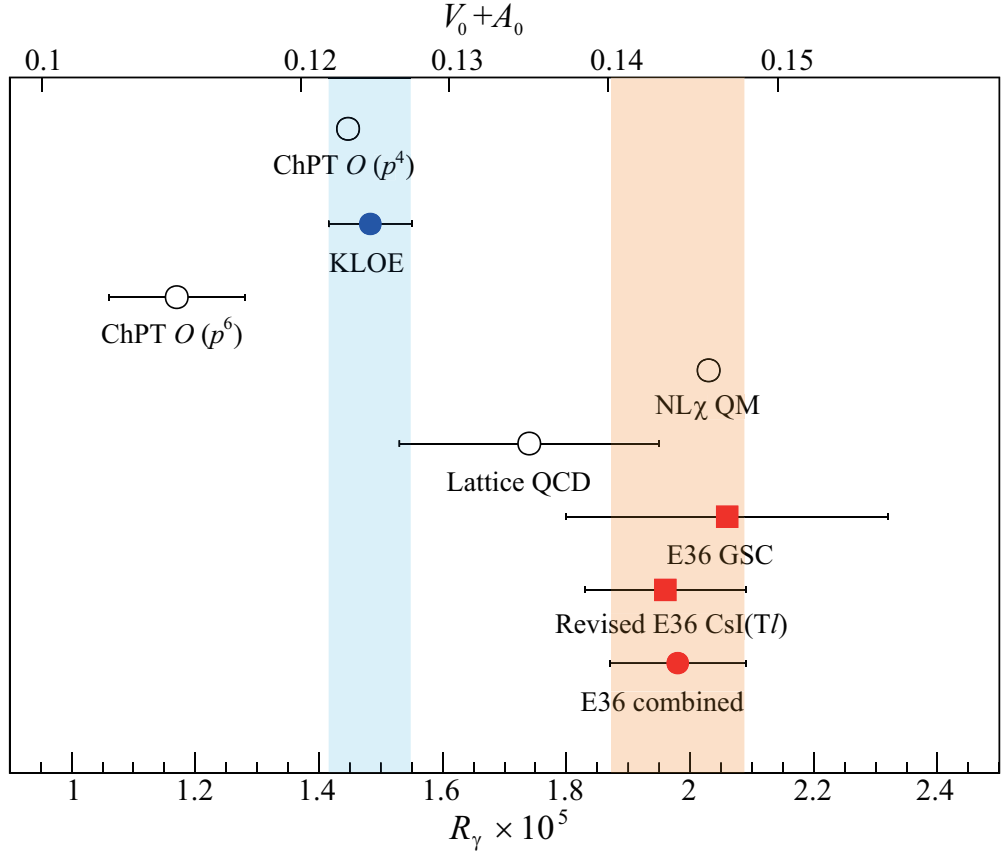}
 \caption{Comparison of the E36 result with other studies. The experimental results (closed symbols) and the theoretical calculations (open symbols) are shown. 
 The NL$\chi$QM~\cite{hosaka1} and ChPT at order $O(p^6)$~\cite{chiPT-2008} results are converted from the form factor ($V_0+A_0$) to the $R_\gamma$ value \color{black} using the values of ($V_0+A_0$) and $R_\gamma$ determined by KLOE~\cite{KLOE2009}. \color{black}
 The combined E36 result based on \color{black} a revision of the CsI(Tl) result from Ref.~\cite{E36_CsI} in this work and the GSC result \color{black} is consistent with the calculations using the lattice QCD and NL$\chi$QM, but 3.9$\sigma$, 4.8$\sigma$, and 5.0$\sigma$ larger than the KLOE result\cite{KLOE2009},  ChPT at order $O(p^4)$~\cite{biji1,biji2,KLOE2009}, and ChPT at order $O(p^6)$, respectively.
 }
 \label{fig.res}
\end{figure}

\section{Result}\label{sec:sec5}
Using the GSC, the $K_{e2\gamma(\gamma)}^{\rm SD}$ branching ratio relative to the $K_{e2(\gamma)}$ decay was determined by adding the total size of the systematic uncertainty as $Br(K_{e2\gamma(\gamma)}^{\rm SD})/Br(K_{e2(\gamma)})=1.25\pm0.14_{\rm{stat}}\pm0.08_{\rm{syst}}$. 
In this paper, the IB process affecting the $K_{e2\gamma}^{\rm SD}$ decay was taken into account for the acceptance calculation in the simulation, which was not the case for the previous E36 $K_{e2\gamma}^{\rm SD}$ analysis using the CsI(Tl) calorimeter~\cite{E36_CsI}.
\color{black} Therefore, the acceptance of the $K_{e2\gamma(\gamma)}^{\rm SD}$ decay was recalculated with the generator of Ref.~\cite{gatti} and applied to the CsI(Tl) analysis. 
After including these IB effects, \color{black} the $Br(K_{e2\gamma(\gamma)}^{\rm SD})/Br(K_{e2(\gamma)})$ value from the CsI(Tl) analysis was revised to $1.19\pm0.07_{\rm{stat}}\pm0.04_{\rm{syst}}$, which supersedes the result of the previous CsI(Tl) analysis~\cite{E36_CsI} and is consistent with this GSC experimental result within the experimental uncertainty. 
Finally, the two results from the CsI(Tl) and GSC analyses were combined in an error-weighted average to be $Br(K_{e2\gamma(\gamma)}^{\rm SD})/Br(K_{e2(\gamma)})=1.20\pm0.07$, \color{black} because the two data sets were totally independent by requiring the CsI(Tl) hit and no-hit in the CsI(Tl) and GSC analyses, respectively. \color{black}
The partial fraction of the $Br(K_{e2\gamma(\gamma)}^{\rm SD})$ in the phase space region ($p>200$ MeV/$c$ and $E_{\gamma}>$ 10 MeV) was then calculated to be $R_{\gamma}=(1.98\pm0.11)\times 10^{-5}$, \color{black} where the systematic effect of this phase space reduction due to the form factor uncertainty used in the analysis was estimated to be negligibly small~\cite{E36_CsI}.
The combined result \color{black} is consistent with the calculation using lattice QCD~\cite{lattice}, but 4.8$\sigma$ and 3.9$\sigma$ larger than ChPT at order $O(p^4)$~\cite{biji1,biji2,KLOE2009} and the KLOE result~\cite{KLOE2009}, respectively.
Also, using the relation of $(V_0+A_0)^2 \propto R_\gamma$ and \color{black} the values of ($V_0+A_0$) and $R_{\gamma}$ determined by KLOE~\cite{KLOE2009},  \color{black} the combined result is consistent with the form factor prediction of  NL$\chi$QM~\cite{hosaka1}, but 5.0$\sigma$ larger than ChPT at order $O(p^6)$~\cite{chiPT-2008}.
The above experimental and theoretical results are summarized in Fig.~\ref{fig.res}.

\section*{Acknowledgements}
We would like to express our gratitude to all member of the J-PARC Accelerator, Cryogenic, and Hadron Experimental Facility groups for their support. 
We also want to thank \mbox{A.R.~Zhitnitsky} for the theoretical support to consider the IB process in the $K_{e2\gamma}^{\rm SD}$ decay.
The present work was supported by JSPS KAKENHI Grant numbers JP26287054, JP15K05113, JP22340059, and JP23654088 in Japan; by NSERC (SAPPJ-2017-00034) and NRC (TRIUMF) in Canada; by Department of  Energy (DOE) DE-SC0003884 and DE-SC0013941 in the United States; and by Russian Science Foundation Grant No. 14-12-00560 in Russia.

%
%

\end{document}